\documentclass[conference]{IEEEtran}
\IEEEoverridecommandlockouts
\usepackage{graphicx}
\usepackage{booktabs}
\usepackage{amsmath,amssymb}
\usepackage{url}
\usepackage{balance}

\newcommand{\sys}{\textsc{RecTile}}
\newcommand{\us}{\,\ensuremath{\mu}s}

\begin{document}

\IEEEpubid{\begin{minipage}[t]{\columnwidth}\footnotesize\raggedright Presented at BAŞARIM 2026, the 9th National High Performance Computing Conference, İstanbul, Türkiye, October 14--16, 2026.\end{minipage}\hspace{\columnsep}\makebox[\columnwidth]{}}

\title{One Structure Is Enough: Record-Native Tiles for\\
Evolving Sparse Matrices on GPUs
}

\author{
Yusuf Mücahit Çetinkaya$^{1,2}$ \\
$^{1}$\textit{School of Informatics, University of Edinburgh,} Edinburgh, UK \\
$^{2}$\textit{Department of Computer Engineering, Middle East Technical University,} Ankara, Türkiye\\
v1ymucah@ed.ac.uk\vspace{-14pt}
}

\maketitle

\begin{abstract}
Sparse matrices that change over time, such as streaming graphs, are usually handled on GPUs with two parts: a packed format such as CSR, ELL, or SELL for computation, and a separate buffer for updates. Periodic maintenance then merges updates, rebuilds data, or changes formats. Our measurements show that this split is often unnecessary. We present RecTile, a tiled storage layer where each tile is kept in one sorted array of 12-byte edge records. The array supports updates and SpMV. For fragmented matrices, with 1 to 21 nonzeros per tile in our workloads, RecTile matches the best format-pinned slab tile at 0.56 to 1.21 times its time. Its tiled layout is also up to 21 times faster than whole-matrix cuSPARSE CSR. Specialized slabs are faster on the patterns they target at medium density, but both approaches converge by 25\% density. RecTile stays within 1.1 to 1.6 times of whole-matrix cuSPARSE COO, our strongest static baseline, while COO cannot be updated in place. RecTile also uses no fixed per-tile device metadata for the stored matrix payload, so its tiles are smaller at realistic occupancies. A fully GPU-based ingest pipeline applies insert, delete, and modify batches at up to 29 million edges per second, 4.5 times faster than the host path. An exact dynamic-programming oracle selects zero folds on every tested workload under its action space and calibrated cost model. Standard folding heuristics use 4 to 15 times more GPU time and up to 4 times more persistent device storage, with no time benefit. End to end, RecTile improves the time-storage tradeoff on all five evolving-graph workloads, including delete-heavy streams. It is also fastest on a dense stress stream with 1220 records per tile. There, folding uses 1.3 times less peak persistent storage but takes 3.9 times more time. RecTile removes the need to maintain a second compute-only representation. For the evolving graph-like, interleaved update/SpMV workloads evaluated here, the mutable record store can serve directly as the compute representation.
\end{abstract}

\begin{IEEEkeywords}
sparse matrices, dynamic graphs, GPU, SpMV, storage formats, streaming
\end{IEEEkeywords}
\section{Introduction}
\label{sec:intro}
Sparse matrix-vector multiplication (SpMV) is widely used in graph analytics and iterative numerical methods. Over the past five decades, many compressed storage formats have been developed for parallel hardware. Examples include CSR, ELL, SELL-$C$-$\sigma$, HYB, and CSR5~\cite{bell2009spmv,kreutzer2014sell,liu2015csr5}. These formats improve memory access and reduce storage costs in different ways. However, they all assume that the matrix does not change. This makes updates expensive. For example, inserting a new nonzero into CSR may require shifting all later entries. Deleting an entry can also leave a gap that CSR cannot directly represent.

Many real workloads do not follow this assumption. Social networks, transaction graphs, and telemetry streams change over time. At the same time, applications may need to recompute rankings, propagate labels, or train models. A common solution comes from log-structured storage~\cite{oneil1996lsm}. It uses a two-tier design. The packed format is kept as the main structure for computation, while new updates are stored in a separate mutable ``delta''. The system periodically merges these updates into the main structure. It may also rebuild the data or choose a different format for some regions. Most GPU systems for dynamic graphs use some form of this design~\cite{green2016custinger,busato2018hornet,winter2018faimgraph,sha2017gpma,king2016dcsr}. A recent VLDB survey studies six such systems~\cite{lin2025dgm}.

Early work on SpMV focused on static matrices. Later work improved memory access, load balance, and storage use. Dynamic graph systems focused more on updates, while LSM-style designs kept updates separate and merged them later. This often leads to a two-tier design with one structure for updates and another for computation. In this paper, we directly measure whether this two-tier design is still useful on modern GPUs and whether one simple mutable representation can store updates and also serve directly as the SpMV representation. Our results show problems with both parts of the design:

\begin{itemize}
\item \emph{Tier 1 (packed format).} A sorted record array is competitive with fixed CSR and SELL tiles in the sparse range seen in evolving graphs. The main difficulty is choosing the right packed format, since CSR and SELL can behave very differently on the same data.

\item \emph{Tier 2 (maintenance).} We use an exact dynamic program to decide when updates should be folded into a packed format. Across the tested workloads, the time-optimal schedule performs no folds. Standard folding policies also add noticeable time and storage cost.
\end{itemize}

Based on these results, we present \sys{}, a record-native tile design. Each $T_s\times T_s$ tile stores one sorted array of 12-byte records. The same array receives updates and is used directly for SpMV. The design does not use folds or format conversions. Internal work such as tail sorting and tombstone compaction remains part of the update path.

Our main contributions are:

\begin{enumerate}
\item \textbf{A record-native tile representation:} Each tile uses one sorted record array for both updates and SpMV. The design avoids a separate packed compute format and keeps per-record storage small.

\item \textbf{A fully on-GPU ingest pipeline:} Updates are grouped, deduplicated, and merged on the GPU. The pipeline reaches up to 29\,M edge updates per second and is $4.5\times$ faster than the host path.

\item \textbf{An exact maintenance schedule:} We use a per-tile dynamic program to choose between deferring updates and folding to CSR or SELL. Across the tested workloads, the optimal time-based schedule performs no folds. To the best of our knowledge, this is the first work to compare such a maintenance policy against an exact optimum.

\item \textbf{An end-to-end evaluation:} We test three real temporal graphs and three synthetic streams against six policies. \sys{} is the fastest method on every workload and improves the time-storage trade-off on the evolving graph workloads.
\end{enumerate}

\section{Background and Related Work}
\label{sec:related}

Sparse matrix-vector multiplication has been studied for many years on GPUs. Bell and Garland examined several common sparse formats, including CSR, ELL, COO, and HYB~\cite{bell2009spmv}. These formats organize nonzero entries in different ways. Their goal is to improve memory access, reduce storage cost, and balance work across GPU threads.

CSR stores nonzeros by row and uses a row-pointer array to locate each row. ELL stores a fixed number of entries for every row. This gives regular memory access, but short rows may require extra padding. HYB combines ELL and COO so that regular parts of the matrix can use ELL while the remaining entries are stored in COO~\cite{bell2009spmv}.

Later work introduced other formats and execution methods. SELL-$C$-$\sigma$ groups rows into smaller slices and reorders rows to reduce padding~\cite{kreutzer2014sell}. CSR5 adds extra information to CSR to improve load balance on different parallel architectures~\cite{liu2015csr5}. Merrill and Garland proposed a merge-based SpMV method that distributes work using the CSR row offsets and nonzero positions~\cite{merrill2016merge}. Their method improves load balance without changing the matrix into a different sparse format.

COO is one of the simplest sparse matrix formats. Each nonzero stores its row index, column index, and value. Bell and Garland also studied GPU COO SpMV and used segmented reduction to combine values that belong to the same row~\cite{bell2009spmv}. COO is easy to build and modify, but it usually stores more index information than CSR.

Tiled methods divide a sparse matrix into smaller two-dimensional regions. This allows different parts of the matrix to be handled separately. TileSpMV uses this approach and selects a suitable representation for each sparse tile~\cite{niu2021tilespmv}. Tiling can improve locality and can also handle different sparsity patterns inside the same matrix. Most of these methods focus on matrices whose structure stays fixed during computation.

Dynamic graph systems focus on data that changes while it is being processed. These systems must support operations such as edge insertion and deletion without rebuilding the complete graph after every update.

cuSTINGER uses dynamic adjacency arrays to support graph updates on GPUs~\cite{green2016custinger}. Hornet uses block-based adjacency storage together with a GPU memory manager~\cite{busato2018hornet}. faimGraph uses paged memory and GPU-side memory management to support fully dynamic graphs under limited memory~\cite{winter2018faimgraph}.

Other systems use more structured layouts. GPMA uses a packed-memory array to keep graph data ordered while leaving free space for later updates~\cite{sha2017gpma}. DCSR extends the CSR idea with segmented storage. A row can use more than one memory segment, which allows new entries to be added without rebuilding the full matrix~\cite{king2016dcsr}.

These methods make different choices about memory use, update speed, and graph processing. Some focus on fast updates, while others focus on keeping the graph compact or ordered. A recent survey by Lin et al.~\cite{lin2025dgm} reviews and evaluates several GPU systems for dynamic graph management.

Dynamic graph storage has also been studied on CPUs. Terrace uses different container types for vertex neighborhoods based on vertex degree~\cite{pandey2021terrace}. Small and large neighborhoods can therefore use different storage methods.

Aspen is built around compressed purely functional trees, called C-trees, for low-latency graph streaming~\cite{dhulipala2019aspen}. Sortledton stores neighborhoods as sorted sets and supports transactional graph updates~\cite{fuchs2022sortledton}. Spruce also targets dynamic graph storage and focuses on high update performance with low memory use~\cite{shi2024spruce}.

Another important line of work comes from log-structured merge trees, or LSM-trees~\cite{oneil1996lsm}. An LSM-tree first writes new data to a mutable structure. The data is later merged into larger and more compact structures. This avoids expensive changes to the main structure for every update, but it introduces extra merge work. LSM-style designs have also been explored for general GPU data structures. GPU LSM applies this idea to a dynamic dictionary on the GPU~\cite{ashkiani2018gpulsm}.

\begin{figure*}[t]
    \centering
    \includegraphics[width=0.9\textwidth]{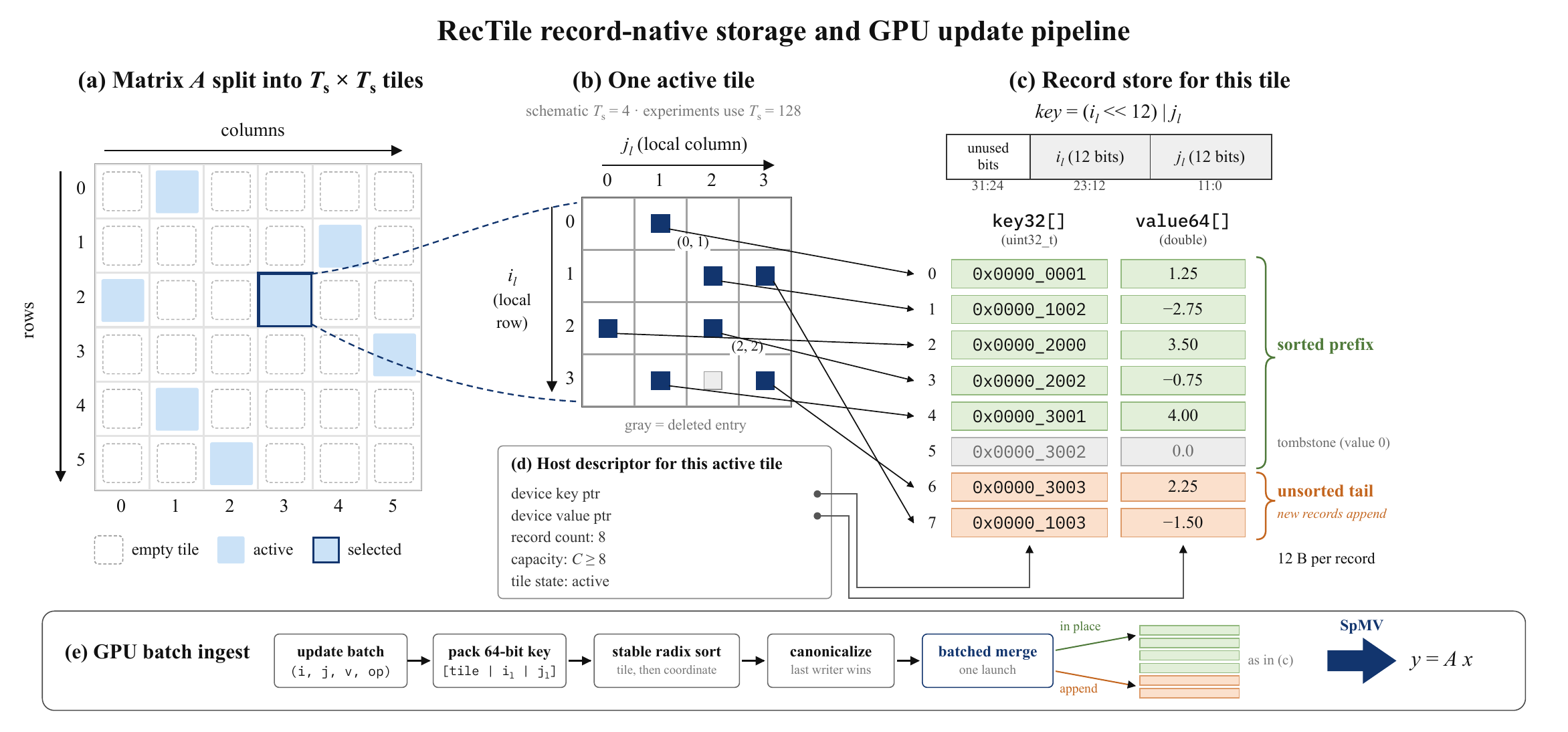}
    \caption{Overview of \sys{}'s record-native tile representation, per-active-tile host metadata, and GPU batch-update pipeline.}
    \label{fig:rectile_overview}
\end{figure*}

\section{\sys{} Design}
\label{sec:design}

\subsection{Tiles and the record store}

As displayed in Figure~\ref{fig:rectile_overview}, \sys{} divides an $m\times n$ matrix into square tiles of size $T_s \times T_s$, where $T_s \le 4096$. The default tile size is 128. Each tile stores its nonzeros as an array of records.

Each record has a 32-bit key and a 64-bit value. The lower 24 bits of the key store the tile-local row and column coordinates: $\mathit{key} = i_\ell \ll 12 \,|\, j_\ell$. Each coordinate uses 12 bits, which limits $T_s$ to 4096. The key and value together use 12 bytes per nonzero. They are stored in a structure-of-arrays layout, with one \texttt{int32} key array and one \texttt{f64} value array. No padding is used.

The record data has no fixed per-tile metadata on the device. Each active tile has a small host-side descriptor, illustrated in Figure~\ref{fig:rectile_overview}, which stores the device pointers, record count, capacity, and tile state. These descriptors are stored only for active tiles, so memory use does not depend on the full size of the tile grid. Section~\ref{sec:metadata} describes the associated SpMV task metadata.

This is different from a CSR tile. With $T_s=128$, a CSR tile uses about 592\,B before storing its first nonzero. This includes 516\,B for row pointers, a 64\,B header, and about 12\,B for its share of the descriptor. Real evolving graph workloads often contain only 1--21 nonzeros per tile. At this occupancy, the fixed storage cost of a packed tile can be much larger than the data stored inside it.

\textbf{Sorted-prefix:} The records in each tile are split into two parts. The first part is sorted by key. The second part is a small unsorted tail. During an update, point lookups use binary search on the sorted prefix and scan only the tail.

The tail is allowed to grow up to $\max(64, n_r/4)$ records. Here, $n_r$ is the total record count in the tile, including tombstones that have not yet been removed. The number of live nonzeros is therefore $n_r$ minus the number of tombstones. When the tail becomes too large, a radix sort over the used 24 key bits sorts the full tile again. Sorting is done as one stream-ordered operation. In normal operation, most tiles remain fully sorted, which is useful for SpMV.

Tail sorting is part of the update path and its cost is included in our measurements. In this paper, ``no maintenance'' means that no folds or format conversions are performed. It does not mean that the record store performs no internal work.

Deleted records remain as tombstones until an explicit compaction step removes them. Compaction uses two \texttt{cub} select operations. The evaluated record-native configuration does not trigger this compaction step. 

\textbf{Semantics:} The store follows the same exact $A = B + D$ semantics as the delta-overlay design. A record in the mutable store is authoritative for its coordinate. Deletions are represented by tombstones with value $0.0$. These tombstones remain in storage until compaction. Their contribution during SpMV is zero.

Update batches may contain insert, delete, and modify operations. The batch is canonicalized before it is merged into the tile store. If the same coordinate appears more than once in a batch, the last operation wins after the stable sort.

Let $r$ be the current record for a coordinate. The update rules are as follows;
\[
\begin{aligned}
(\mathrm{I},\varnothing) &\to \text{new record } v, &
(\mathrm{I},r) &\to r=v, \\
(\mathrm{D},\varnothing) &\to \text{no-op}, &
(\mathrm{D},r) &\to r=0, \\
(\mathrm{M},\varnothing) &\to \text{no-op}, &
(\mathrm{M},r) &\to r=v.
\end{aligned}
\]

A deleted record is therefore kept in place with value zero. If an entry in a base slab is replaced by a record, the old slab entry is also set to zero. This keeps the $A=B+D$ state correct.

Record capacity grows geometrically. When more space is needed, the capacity is doubled.

Explicit zero entries are not stored. Inserting a value of $0.0$ is treated as a delete. We use the term ``exact'' for the stored matrix state. After every update batch, the stored $(i,j,v)$ entries must match the reference implementation. SpMV results are checked using fp64 round-off tolerance because different layouts may sum values in a different order. The reference implementation applies every operation in the test suite.

\subsection{SpMV over records}

The SpMV kernel processes the records of each tile in groups of 32. Each group is assigned to one warp. A 256-thread block therefore contains eight warps.

The sorted prefix keeps records in non-decreasing row order. This allows each warp to combine values from the same row before writing the result. The warp performs an inclusive segmented scan using register shuffle operations. The row index is used to mark segment boundaries.

After the scan, the last lane of each row group performs one \texttt{atomicAdd}. If a row crosses a warp boundary, each warp writes one partial result. This case is valid and occurs rarely. Records in the unsorted tail cannot use the same grouping, so they use one atomic operation per record.

The segmented reduction improves performance when tiles become dense. At 25\% density, it is $1.5$--$1.7\times$ faster than using one atomic operation per record. At low occupancy, both versions have similar behavior because each tile contains only a small number of records.

\subsection{On-GPU ingest}
\label{sec:design-ingest}

An update batch is uploaded once as raw $(i,j,v,\mathit{op})$ records. A packing kernel converts each update into a 64-bit key:

\[
[\,\mathit{tile}:40\,|\,i_\ell:12\,|\,j_\ell:12\,].
\]

The GPU then performs one stable radix sort using CUB library. The sort first groups updates by tile. It also places updates for the same coordinate in arrival order.

A run-length encoding pass finds groups of updates that refer to the same coordinate. A fold kernel then applies the update rules to each group. This produces one canonical result for each coordinate. A second run-length encoding pass builds the list of update runs for each tile.

The canonical updates are then merged into the touched tile stores with one batched kernel launch. The merge searches the sorted prefix of each tile. Existing records are updated in place, while new records are appended. Large tiles are split into chunked block tasks so that the GPU can process several parts of a tile in parallel.

The host performs only range checks, per-tile memory allocation, and statistics collection. Two short synchronizations are used for each batch to read the data-dependent run information from the GPU.

We also keep the older host ingest path for comparison. That path uses host-side hash-map bucketing and separate sorting for each tile.

\subsection{Concurrency and correctness machinery}

Updates, SpMV, and baseline rebuilds use separate non-blocking CUDA streams. CUDA events control the order between these streams. A reader cannot access a tile while an update is only partly applied.

A rebuild may also become outdated if a newer update reaches the same data before the rebuild is activated. In this case, the rebuild result is discarded and is never made visible to readers.

Correctness is checked with 134 GoogleTest suites and 79 Python tests. The tests include randomized orders of updates and SpMV operations. Stored matrix states are compared with a bit-exact reference implementation.

\subsection{Metadata and task discovery}
\label{sec:metadata}

GPU kernels need a way to find the records that belong to each tile. Each active tile has one sparse host descriptor. The descriptor stores device pointers, record counts, capacity, and tile state. The GPU also uses task arrays for SpMV. Each active tile has one 48\,B queue descriptor. It contains the tile origin, key and value pointers, record count, and sorted-prefix length. There is also one 8\,B warp task for every 32 records.

The task arrays are rebuilt only when the set of active tiles changes. They are not rebuilt before every SpMV call. Rebuilding the arrays requires a host scan, one host-to-device copy, and one synchronization. This work is outside the GPU-event timing but is included in the reported wall-clock time.

On sx-mathoverflow, the record-native task arrays use about 4.0\,MB for 78k tiles and 1.3M records. The persistent record storage uses 15.8\,MB. Task arrays are not included in the persistent storage metric for any method.

The statement that record tiles use no fixed per-tile device metadata therefore refers only to the stored matrix payload. The task system still uses 48\,B per active tile and 0.25\,B per record for every representation.

\subsection{Fair baselines}
\label{sec:baselines}

We strengthen the slab baselines before comparison. The grouped CSR kernel skips empty rows using warp ballot compaction, which improves performance by $4$--$29\times$ on fragmented tiles. Baseline folds use a batched rebuild engine that reduces the per-tile rebuild cost from about $200\us{}$ to 11.6\us{} for large batches. Fold policies also use a GPU-calibrated cost model, hysteresis, and a byte-aware memory guard.

The full C++17/CUDA implementation, the Python test and experiment code, and the drivers used to produce the reported results are shared on a public Github repository \footnote{\url{https://github.com/tweetpie/RecTile}}.

\section{The Format}
\label{sec:crossover}

\subsection{Methodology}

We compare the same tile contents using six SpMV configurations. These are records with segmented reduction, records without segmented reduction, fixed CSR slabs, fixed SELL-32 slabs, whole-matrix cuSPARSE CSR, and whole-matrix cuSPARSE COO~\cite{cusparse}. SELL-32 uses a slice height of 32, sorts rows by length, and skips padding entries. Each experiment uses a grid of 4096 tiles with the same fill pattern. With $T_s=128$, the full matrix has about 524k rows. We measure 50 SpMV iterations with CUDA events after a warm-up period. All vectors remain on the GPU, and no host-device transfers are included in the measured region. We test occupancies from 1 to 4096 records per tile, which covers densities up to 25\%. We use uniform, single-row, and banded fill patterns. We test tile sizes of 128 and 256 on both GPUs.

\begin{figure}[h]
\centering
\includegraphics[width=0.9\linewidth]{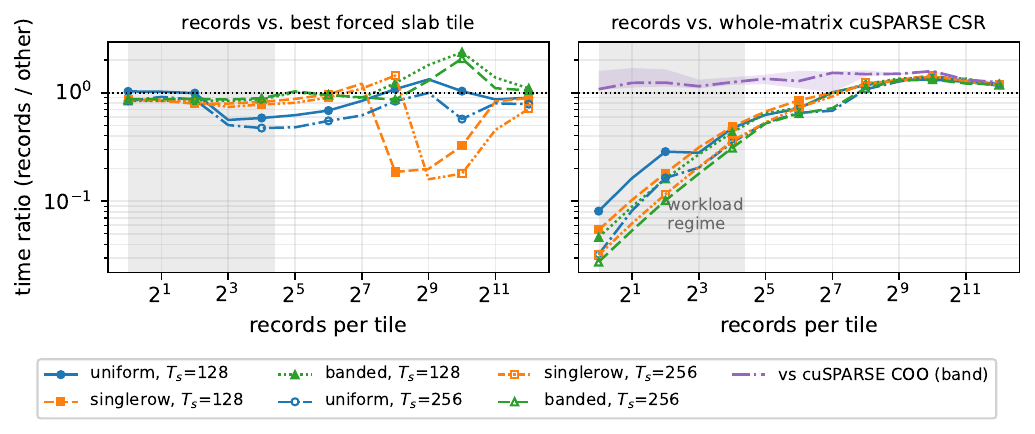}
\caption{SpMV time of the record format relative to the best fixed slab format on the left and whole-matrix cuSPARSE CSR on the right. Values below 1.0 mean that records are faster. The purple band shows the comparison with whole-matrix cuSPARSE COO.}
\label{fig:crossover}
\end{figure}

\subsection{Results}

Fig.~\ref{fig:crossover} and Table~\ref{tab:crossover} show how performance changes with tile occupancy. In the fragmented range of at most 128 records per tile, records take $0.56$--$1.21\times$ the time of the best fixed slab format across the tested patterns and tile sizes. On Blackwell, the range is $0.56$--$1.23\times$. This occupancy range covers all of our evolving graph workloads. The tiled record layout is also up to $21\times$ faster than whole-matrix cuSPARSE CSR in this range. cuSPARSE CSR still processes the row-pointer structure for about 524k rows, even when most rows contain very little data.

The difference between CSR and SELL is much larger than the difference between records and the best slab format. CSR and a fixed SELL tile can differ by up to $10\times$ on the same tile contents. The faster format depends on the fill pattern. For example, with a single-row pattern, SELL takes $0.234$\,ms while CSR takes $0.012$\,ms. The tested selection rule uses SELL only when the mean row length is at least 2. This rule chooses the slower format across part of the tested range.

At medium occupancy, from 256 to 1024 records per tile, the best format depends more strongly on the fill pattern. SELL performs best on banded tiles and can be up to $2.4\times$ faster than records. Records are $3$--$5\times$ faster than SELL on single-row tiles. For uniform tiles, the difference stays within about $\pm35\%$.

At high occupancy, from 2048 to 4096 records per tile, which corresponds to about 12--25\% density, the formats become much closer. Records take $0.83$--$1.38\times$ the time of the best slab format and $1.14$--$1.45\times$ the time of whole-matrix cuSPARSE CSR. We observe the same general behavior on both GPUs and with both tile sizes.

Whole-matrix sorted COO is the strongest static baseline across the tested occupancies. On Ada, cuSPARSE COO is $1.1$--$1.6\times$ faster than the tiled record pass. Both use a sorted record-style layout, but cuSPARSE COO does not divide the matrix into tiles. The tiled layout gives up some static SpMV speed in return for local updates. An update to the tiled structure only needs to touch the affected tiles, while updating a whole-matrix COO layout requires changing the global structure. Static SELL can also be faster when the fill pattern matches its layout, especially in the medium-density band.

\begin{table}[t]
\caption{SpMV time per pass on the 4096-tile grid with $T_s=128$ on RTX 6000 Ada. Bold marks the fastest result in each row.}
\label{tab:crossover}
\centering
\footnotesize
\setlength{\tabcolsep}{3.6pt}%
\begin{tabular}{@{}llrrrrr@{}}
\toprule
Occ. & Pattern & records & CSR-t & SELL-t & cuSP-CSR & cuSP-COO \\
\midrule
16   & uniform   & 0.010 & 0.030 & 0.017 & 0.021 & \textbf{0.008} \\
128  & uniform   & 0.015 & 0.110 & 0.018 & 0.015 & \textbf{0.010} \\
1024 & uniform   & 0.056 & 0.172 & 0.054 & 0.043 & \textbf{0.035} \\
1024 & banded    & 0.055 & 0.171 & \textbf{0.024} & 0.040 & 0.035 \\
4096 & uniform   & 0.288 & 0.322 & 0.321 & 0.249 & \textbf{0.236} \\
\bottomrule
\end{tabular}
\end{table}

We also tested dynamic switching between CSR and SELL. The performance gap between the two formats can reach $10\times$, but the faster format changes with the fill pattern. Converting one tile between formats costs about $200\us{}$ when done alone and 11.6\us{} per tile when conversions are batched. The conversion cost requires about $10^3$--$10^5$ later uses of the tile before it can be recovered.

We also measured the storage cost of folding records into slab formats. On the real graph workloads, folding increased peak storage by $2.7$--$4\times$. At low tile occupancy, the fixed metadata of each slab takes a large amount of space compared with the small number of records stored in the tile.

\section{Maintenance Scheduling}
\label{sec:oracle}

\subsection{The maintenance oracle}

For each tile, we use a dynamic program over windows $w=1..W$. The state is defined by the current storage format and the size of the delta bucket. At each window, the scheduler can choose one of three actions: defer the updates, fold them into CSR, or fold them into SELL.

Let the state at the start of window $w$ be $s=(f,b)$, where $f$ is the current format and $b$ is the delta-size bucket. The update volume for tile $t$ in window $w$ is $g_{t,w}$. We use several cost functions measured on the GPU. $U(\cdot)$ is the cost of keeping an unfolded store for one window. $S_f(\cdot)$ is the SpMV cost of format $f$. $D(\cdot)$ is the cost of processing the record store. $F(\cdot)$ is the cost of folding the tile. $K_w$ is the number of SpMV passes in window $w$. We use $K_w=10$ unless stated otherwise.

\begin{align}
&b' = \min(b + g_{t,w}, B{-}1),\quad
c_w = U(b) + K_w[S_f + D(b')] \nonumber\\
&V_w(f,b) = c_w + \min\big( V_{w+1}(f,b'),
{\textstyle\min_{g}} F(b') {+} V_{w+1}(g,0) \big)\!,
\label{eq:dp}
\end{align}

The equation compares two choices. The first choice is to defer the updates and keep the larger delta. The second choice is to fold the tile into CSR or SELL, pay the fold cost, and reset the delta bucket to zero. The objective is total GPU execution time. Memory use is not part of the objective or a constraint.

All cost values are calibrated on the GPU. The dynamic program checks every allowed action at each state and chooses the one with the lowest total future cost. We solve it backward from the last window to the first. Tiles are independent in the model, so each tile can be solved separately and the results can be added together.

We also checked the implementation with brute-force enumeration on small cases. This included 25 random traces and several adversarial cases. The vectorized implementation can solve 525{,}751 tiles over 100 windows in a few seconds. The result is exact only for the actions and cost model defined above. We do not claim it to be optimal over every possible maintenance method or objective.

\subsection{Effect of fold cost}
\label{sec:oracle-pricing}

The result mainly depends on two measured properties. At the observed occupancies, SpMV over records does not cost more than the slab pass, and the cost of keeping an unfolded store does not increase with its size. Folding therefore adds a positive cost without reducing future GPU time in the model. For the fragmented workloads, the scheduler chooses no folds for any positive fold cost. On the dense stream, it still chooses no folds when the fold cost is reduced to 1\% of its calibrated value.

The calibrated fold cost is also favorable to folding. It is close to the lowest cost reached by the batched rebuild engine. We replay the selected schedule in the full system and observe about 12\% prediction error on the graph workloads. The scheduler still chooses no folds.

\subsection{Sensitivity analysis}
\label{sec:oracle-sensitivity}

We also change the two main fitted costs and solve the dynamic program again. On sx-mathoverflow, folding appears only after the carrying cost or the assumed slab SpMV advantage is increased far beyond the measured values. The dense stream is more sensitive because it has about 1220 records per tile. We also vary the number of delta buckets from 2 to 16 and the SpMV reuse factor from 10 to 200. The scheduler still chooses no folds in both workloads.

\subsection{Scheduling results}
\label{sec:oracle-verdict}

The time-optimal schedule chooses zero folds and zero format conversions on every tested workload. This covers between 78k and 612k active tiles, and runtime counters confirm that no folds are performed. Replaying the oracle schedule gives the same runtime as the no-maintenance configuration within measurement noise.

The reason is simple in the fitted model. Keeping records does not become more expensive as the delta grows, and SpMV over records is not slower in the tested fragmented range. Folding therefore adds cost without saving enough GPU time later.

This result only considers GPU execution time. A system with a strict memory limit may still choose to fold in order to reduce storage. Conventional fold heuristics perform between 114k and 598k folds per run in our experiments. They use $4$--$15\times$ more GPU time and up to $4\times$ more storage.
\section{Ingest Performance}
\label{sec:ingest}

We compare the host ingest path with the on-GPU ingest pipeline using batch sizes from $10^5$ to $10^7$. The tests use a $524\text{k} \times 524\text{k}$ matrix with 5M initial nonzeros. We test two update patterns. In the first pattern, updates are concentrated on a few thousand tiles, which is closer to real workloads. In the second pattern, every update goes to a different tile in a grid with 16.7M tiles.

When updates have realistic locality, the GPU pipeline reaches \textbf{29.0\,M edges/s} for a batch of 10M updates. The host path reaches 6.4\,M edges/s, so the GPU pipeline is $4.5\times$ faster. Its throughput also continues to improve as the batch size grows. When updates are spread across different tiles, both methods slow down. The GPU pipeline reaches 2.9\,M edges/s and the host path reaches 0.8\,M edges/s. Most of the remaining cost comes from host-side allocation when a new tile is created.

The GPU pipeline uses a fixed sequence of steps for every batch. It packs the update keys, performs one stable radix sort, runs two run-length encoding passes, applies the update rules, and finishes with one batched merge. It also uses two short host synchronizations per batch. The number of kernel launches does not depend on how many tiles are touched.

Published systems can report higher update rates under different update rules. For example, faimGraph reports up to 200\,M edges/s~\cite{winter2018faimgraph}. That number measures adjacency updates and does not include the same deduplication, delete and modify handling, or exact matrix-state checks used here. We therefore use it only as a reference point and not as a direct baseline.

Several implementation changes also improved the ingest path before moving the main work to the GPU. Using one merge launch per batch avoids launching work separately for each touched tile and gives up to a $7.7\times$ improvement when updates are widely spread. Appending into the sorted-prefix structure also keeps update cost nearly independent of the current store size and gives up to an $8\times$ improvement at 78k records. Applying all of these changes reduced update time by about $2$--$3\times$ before the full GPU pipeline was added.

\begin{table*}[t]
\caption{Median GPU time and peak persistent device storage relative to \sys{}. $K=10$, $T_s=128$, 80--100 windows, and 3 runs per configuration.}
\label{tab:campaign}
\centering
\footnotesize
\begin{tabular}{@{}lrrrrrr@{}}
\toprule
 & \multicolumn{2}{c}{\sys{} (absolute)} &
 \multicolumn{2}{c}{Conventional vs.\ \sys{}} &
 \multicolumn{2}{c}{Best hybrid vs.\ \sys{}} \\
\cmidrule(lr){2-3}\cmidrule(lr){4-5}\cmidrule(lr){6-7}
Workload & GPU (ms) & Stor.\ (MB) & time & storage & time & storage \\
\midrule
sx-mathoverflow (88.6k nodes) & 244.5 & 15.8 & $7.8$--$9.0\times$ & $4.0\times$ & $1.09\times$ & $1.27\times$ \\
sx-superuser (567k nodes) & 1641.4 & 118.3 & $6.0$--$6.2\times$ & $3.9\times$ & $1.12\times$ & $1.42\times$ \\
email-Eu-core (1.0k nodes) & 28.5 & 0.4 & $4.6$--$5.1\times$ & $1.0\times$ & $1.54\times$ & $1.04\times$ \\
synthetic, insert-only & 573.6 & 99.2 & $11.1$--$14.6\times$ & $3.1\times$ & $1.50\times$ & $3.10\times$ \\
synthetic, delete-heavy & 278.1 & 95.6 & $4.2$--$5.0\times$ & $3.2\times$ & $2.14\times$ & $3.18\times$ \\
synthetic, dense (1220 rec/tile) & 792.8 & 543.5 & $3.9\times$ & $0.76\times$ & $1.21\times$ & $1.07\times$ \\
\bottomrule
\end{tabular}
\end{table*}

\section{End-to-End Evaluation}
\label{sec:eval}

\subsection{Setup}

We run the experiments on a dual-socket Xeon Gold 6326 server with two GPUs: an NVIDIA RTX 6000 Ada with 48\,GB of memory and an RTX PRO 6000 Blackwell with 96\,GB, with CUDA 13.2. We use three temporal graphs from SNAP~\cite{snapnets}: sx-mathoverflow, sx-superuser, and email-Eu-core. We replay their edges in timestamp order. The first 30\% of the edges are used to build the initial matrix, and the remaining 70\% are streamed as inserts. We also use three synthetic workloads. The insert-only and delete-heavy workloads start from a $100\text{k} \times 100\text{k}$ matrix with 1M nonzeros. The delete-heavy workload uses a 20/60/20 mix of insert, delete, and modify operations. The dense workload uses a $16.4\text{k} \times 16.4\text{k}$ matrix with 20M nonzeros and about 1220 records per tile. This workload places the tiles in a density range where packed formats are expected to work well.

Each configuration is normally run three times, and we report the median. The $K=200$ experiments use two runs. For the closest comparisons with the best hybrid, we use 12 runs. Across the 42 workload and policy combinations, the median coefficient of variation for GPU time is 0.9\%. The largest variation is seen on email-Eu-core, where it reaches 25\%. In those cases, the difference between the compared methods is still at least $1.5\times$. GPU costs are measured with CUDA events. 

\textbf{Storage:} We define peak persistent device storage as the largest amount of allocated GPU memory used by the stored matrix during any window. For slab tiles, this includes the packed arrays and about 592\,B of metadata per tile. For record tiles, it includes $12\,\mathrm{B} \times \mathrm{capacity}$, including unused capacity and unreclaimed tombstones. Host descriptors and temporary buffers are excluded for all methods. On sx-mathoverflow, this gives 27.3\,MB for record-native storage, 34.4\,MB for the best hybrid, and 108.2\,MB for fold-always. Folding methods also need temporary memory for rebuilds.

\textbf{Workload:} The three temporal graphs contain 506k, 1.44M, and 332k timestamped edges for sx-mathoverflow, sx-superuser, and email-Eu-core. After the first 30\% is used to build the initial matrix, the remaining updates arrive in fixed-size windows. The average window contains about 3.5k updates for sx-mathoverflow, 10k for sx-superuser, and 2.3k for email-Eu-core. With $T_s=128$, a typical window touches 6.2--18\% of the active tiles. Updates are not evenly distributed across tiles, with a cross-tile coefficient of variation between 3.3 and 4.7. The tiles are also very sparse, with only 1--21 nonzeros per tile.

We compare seven policies. \emph{record-native} is \sys{}. \emph{never} starts with a slab base and keeps later updates as records without folding them back into the slab. This is the strongest hybrid configuration. \emph{always} folds every touched tile after each window. \emph{threshold} folds a tile when its record count reaches 25\% of the slab size. \emph{scheduler} is the full adaptive system with format selection and folding rules. \emph{scheduler-pure} keeps the calibrated folding policy but removes other adaptive choices. \emph{oracle-replay} follows the schedule produced by the dynamic program.

\subsection{Results}

Table~\ref{tab:campaign} shows the end-to-end results. \sys{} has the lowest GPU time on every tested workload. On the evolving graph workloads, it also uses less persistent storage than the other policies. On the dense workload, folding uses less storage but takes more GPU time.

Compared with the conventional fold-always and threshold policies, \sys{} uses $4.2$--$14.6\times$ less GPU time on the evolving workloads. It also uses about $3$--$4\times$ less persistent storage on most of these workloads.

The comparison with the never-fold hybrid is closer. \sys{} is $1.07$--$2.14\times$ faster and uses $1.07$--$3.2\times$ less storage. The smallest time differences appear on sx-mathoverflow and sx-superuser. We repeat these cases 12 times. The median time ratios are $1.09\times$ and $1.07\times$. Their bootstrap 95\% confidence intervals are $[1.08,1.11]$ and $[1.05,1.12]$. The delete-heavy workload shows the largest gap, where the hybrid takes $2.14\times$ more GPU time. Record tombstones can remain in place and are skipped during SpMV, while dead slab entries remain until a rebuild.

The dense workload has about 1220 records per tile. Here, folding reduces peak storage from 544\,MB to 411\,MB, about $1.3\times$. This comes at a GPU-time cost of $3.9\times$. \sys{} is still faster than the best never-fold hybrid and also uses slightly less storage than that hybrid.

The same ordering is seen on both Ada and Blackwell. Changing the reuse factor to $K \in \{50,200\}$ and changing the tile size to $T_s \in \{256,512\}$ also keeps the same ordering.

The full adaptive scheduler performs poorly on several workloads. It uses format selection, lifecycle rules, and threshold-based folding. These operations work as intended but they add work that does not recover enough cost later. On some workloads, the scheduler takes $18$--$26\times$ more GPU time and uses up to $8\times$ more storage than \sys{}.

\subsection{Ablations}
\label{sec:ablation}

\textbf{1. Segmented reduction:} We disable segmented reduction while keeping the same record layout and matrix contents. At 4096 records per tile, segmented reduction improves SpMV by $1.5$--$1.7\times$. With $T_s=256$, the improvement is about $1.2\times$. At 2048 records per tile, the difference is below 10\%. At lower occupancies, the two versions perform almost the same. The main benefit of segmented reduction therefore appears at higher tile densities.

\textbf{2. Record width:} We also compare the 12\,B record format with a 16\,B version that stores the local row and column as separate 16-bit values. Packing both coordinates into one 32-bit key reduces peak storage by $1.9$--$2.1\times$ on the real graph workloads and removes two kernel passes. The 16\,B version still performs well in time but CSR becomes smaller once occupancy grows above about 150 records per tile. With the packed 12\,B records, the record format remains smaller at the occupancies seen in the graph workloads.

\textbf{3. Sensitivity:} We vary the SpMV reuse factor, tile size, GPU architecture, and tile density. The ordering between the methods does not change, and the oracle continues to choose no folds. On sx-mathoverflow, the conventional policy is $7.8\times$ slower at $K=10$, $4.3\times$ slower at $K=50$, and $5.4\times$ slower at $K=200$. With $T_s=256$ and $T_s=512$, the penalties are $6.9\times$ and $5.1\times$. On Blackwell, the penalty is $7.5\times$. For sx-superuser at $K=50$, the conventional policy is $5.6\times$ slower. On the dense workload, it is $3.9\times$ slower. Larger reuse factors and larger tiles can reduce the gap but the effect is not always monotonic because folded hybrid tiles also pay their SpMV cost more often as $K$ increases.

\textbf{4. Long-running churn:} We also test a stream that keeps about 1M live nonzeros throughout the run. Each window deletes 4k existing entries, inserts 4k new entries, and modifies 2k entries. The test lasts 600 windows, which is six times longer than the main experiments. Without tombstone removal, the number of stored records grows from 1.0M to 3.4M, and 71\% of them are dead by the end. Allocated capacity grows by 24\%, and SpMV time increases by 18\%. Periodic folding keeps the number of record tombstones low, but it takes $3.3\times$ more time and uses about $4\times$ more storage because the sparse slabs repeatedly pay their fixed metadata cost. A better option for this case is a record compaction pass triggered by the fraction of dead records.

\section{Discussion and Limitations}

\sys{} benefits from three properties of the evaluated regime. First, at low tile occupancy, packed local coordinates are inexpensive while CSR's fixed row-pointer and metadata costs are poorly amortized. Second, segmented reduction recovers much of the row-structured SpMV benefit without requiring a separate representation. Third, the CSR/SELL winner varies with fill pattern, so maintaining a packed representation is useful only when its future compute savings recover conversion cost; this does not occur in our measured workloads.

The conclusions are scoped to interleaved update/SpMV workloads with tile densities up to 25\%. \sys{} currently uses fp64 values and $T_s \leq 4096$; fp32, multi-GPU operation, and non-graph workloads remain unevaluated. Highly dispersed updates remain limited by host-side tile allocation, and sustained delete-heavy workloads require tombstone compaction. The scheduler optimizes GPU execution time only; memory-constrained or latency-constrained objectives may rationally choose different maintenance schedules. Static matrices and regular medium-density patterns remain better suited to tuned libraries such as cuSPARSE or SELL.

\section{Conclusion}

\sys{} uses one tile-local record representation for both updates and SpMV, avoiding the packed compute representation and periodic folding used by conventional two-tier designs. Across the evaluated evolving graph-like workloads, it provides the best end-to-end GPU time and improves the time-storage trade-off; an exact dynamic program under the calibrated cost model selects no folds. Even on the dense stress workload, folding reduces storage only by trading away execution time. For the evaluated interleaved update/SpMV workloads with tile densities up to 25\%, our results show that mutable record storage can serve as a first-class compute representation and that maintenance policies should be justified against the cost of doing nothing.

\section*{Acknowledgment}
We would like to acknowledge the use of the computation resources supported  by METU under the grant no. ADEP-312-2024-11484.

\balance

\bibliographystyle{IEEEtran}
\bibliography{refs}

\end{document}